# What out-of-the-box LLMs can(t) do in law? A Turing test in Italian exams for lawyers, judges and notaries

Germana Bertoli*; Ilaria Amelia Caggiano**; Francesca Lagioia***; Riccardo Rovatti****; Giovanni Sartor†; Emiliano Troisi‡

**Summary.**

**Abstract.**
The article reports on a blind “Turing Test” experiment, assessing the performance of out-of-the-box leading LLMs on three Italian legal professional exams: the Bar, Judges and Notary exams. Leading LLMs were asked to generate full written exam papers, which were made indistinguishable from human submissions and anonymously evaluated by expert examiners, using the same criteria applied in real examinations. Results reveal marked differences across both models and tasks. While some LLMs match or exceed top human performance in adversarial legal argumentation and doctrinal analysis, all models fail in the notary exam, which requires goal-directed legal planning under strict formal and substantive constraints. Beyond ranking models, the study identifies task-specific strengths, limitations and recurring legal failure patterns. Although limited to out-of-the-box systems, the findings provide qualitative evidence on the current scope and boundaries of LLMs’ legal competence across distinct professional tasks.

## I. Introduction

The rapid evolution of Large Language Models (LLMs) has sparked a debate on the capacities of AI, particularly concerning its ability to replicate human professional expertise. While recent advancements show that frontier models can achieve near-human, or even “superhuman”, performances on tasks pertaining to STEM domains[1] – e.g., mathematical and physical problem-solving, coding and logical theorem proving – the legal domain provides distinct challenges, so far explored to a limited extent [2].

* Attorney at Law, Turin Bar. Coordinator of the Artificial Intelligence Commission of the Turin Bar Association. Email: germana.bertoli@studiolegalebertoli.it
** Full Professor of Private Law, Department of Legal Sciences, University of Naples Suor Orsola Benincasa. Deputy director of the Research Centre of European Private Law (ReCEPL) at University of Naples Suor Orsola Benincasa. Email: ilaria.caggiano@unisob.na.it
*** Associate Professor of Legal Informatics, AI and Law at Department of Law and CIRSFID-Alma AI, University of Bologna. Email: francesca.lagioia@unibo.it
**** Full Professor of Electronics, Department of Electrical, Electronic, and Information Engineering “Guglielmo Marconi”, University of Bologna. Email: riccardo.rovatti@unibo.it
† Full Professor of Legal Informatics and Philosophy of Law at Department of Law and CIRSFID-Alma AI, University of Bologna and Professor of Legal informatics and Legal Theory at the European University Institute (EUI). Email: giovanni.sartor@unibo.it
‡ PhD in Private Law, University of Naples Suor Orsola Benincasa. Senior Researcher at the Research Centre of European Private Law (ReCEPL), University of Naples Suor Orsola Benincasa. Email: emiliano.troisi@unisob.na.it
[1] K. Feng et al, ‘PHYSICS: Benchmarking Foundation Models on University-Level Physics Problem Solving’ arXiv preprint arXiv:2503.21821 (2025); C. Liu et al, ‘Fimo: A Challenge Formal Dataset for Automated Theorem Proving’ arXiv preprint arXiv:2309.04295 (2023); X. Zhang et al, ‘PhysReason: A Comprehensive Benchmark Towards Physics-Based Reasoning’ arXiv preprint arXiv:2502.12054 (2025); L. Zheng et al, ‘A Reasoning-Focused Legal Retrieval Benchmark’, in *Proceedings of the 2025 Symposium on Computer Science and Law* (2025), 169-193.

Unlike STEM domains, where rules are often absolute, context-independent and benchmarks allow for straightforward validation – e.g., by checking final numerical answers or employing formal verifiers – legal reasoning is inherently tied to jurisdiction-specific norms, linguistic nuances, and the ability to apply abstract principles to complex real cases. Indeed, legal reasoning remains a critical and distinct frontier for AI, requiring specialized domain knowledge and advanced cognitive abilities as needed for the interpretation of precedents, statutory analysis, and complex legal inference. This distinction is crucial for understanding why current benchmarking strategies are often insufficient. In particular, there is a persistent gap between "rule recall" and "rule application"[3]. While LLMs excel at retrieving legal information or answering multiple-choice questions – that mimic the pattern matching observed in STEM fields – they notably struggle with long-form legal reasoning that requires a structured, multi-step application of rules, precedents and principles to open-ended problems. In law, the challenge is not just finding correct answers, but building a coherent, consistent, and legally sound reasoning that may pass professional scrutiny. As LLMs are increasingly deployed for high-stakes legal tasks – including judgment prediction, summarization, case retrieval, drafting of contracts and legal opinions and decisions, – it is imperative to thoroughly examine their capabilities in these different tasks. A failure to do so could lead to serious real-world consequences, in a domain where accuracy and reliability are non-negotiable.

Our research aims at testing the ability of leading LLMs – Claude 4 Opus, GPT-5, DeepSeek R1, and Gemini 2.5 Pro – to pass the Italian exams for attorneys (bar exam), judges, and notaries. Each exam requires distinct cognitive capacities, encompassing doctrinal understanding, legal reasoning, and task-specific application. As further detailed in section 2, existing benchmarks primarily focus on university-level examinations or standardized bar exams, mostly in common law jurisdictions (which often knowledge of precedents or multiple-choice accuracy). We shift the focus to the highest tiers of professional qualification in the Italian civil law system. These exams represent an ideal "stress test" for LLMs, since they move beyond the simple recall of legal knowledge into the realm of targeted doctrinal analysis, strategic argumentation and goal-directed planning. Moreover, the diversity of legal professional exams provides a suitable proxy for assessing how and to what extent LLMs acquire and deploy legal knowledge, across different tasks and levels of complexity.

The paper is organized as follows. Section 3 provides an overview of the Italian legal professional exams. Section 2 reviews related works, positioning our study within the current landscape of legal AI benchmarking. Section 4 describes the experimental setting, the models used and the adopted methodology. In section 5, 6 and 7, we detail the evaluation criteria, the applied scoring grids and present the results, providing a comparative analysis across models, as well as a discussion of the identified legal failures. Section 8 concludes on the implications of our findings relatively to AI in legal professions and discusses future research directions.

## II. Background and related works

Research on LLMs legal competences has grown rapidly. An increasing number of works primarily focus on (i) the performance analyses of LLMs on legal exams and (ii) the design of benchmarks, that approximate exam questions or doctrinal reasoning tasks.

The evaluation of LLMs through professional and academic examinations has become a standard benchmark for assessing "General Artificial Intelligence". However, recent literature has shifted from general benchmarks to domain-specific and jurisdiction-specific evaluations. Initial milestones established that frontier models could perform on par with human candidates in standardized environments. Significant assessments include Katz et al., 2024[4] and Martinez, 2025[5] on the US bar exam, Freitas & Gomes, 2023[6] on the Brazilian Bar exam, and Juvekar, 2025[7] for India. More recently, an analysis of LLMs' performance in a wide range of multiple choice and open questions in law school exams has been proposed by LEXam[8], a massive benchmark derived from 340 exams across 116 courses, revealing that while LLMs excel at multiple-choice formats, they struggle with open-ended questions requiring structured, multi-step reasoning. At the same time international benchmarks for legal reasoning have been established: LegalBench[9], LawBench[10], LexEval[11], LEXTREME[12] or LexGLUE[13]. Similarly, the OAB-Bench[14] provides a framework for evaluating legal writing in the Brazilian Bar Exam, revealing that even frontier models like GPT-4o and DeepSeek-R1 exhibit inconsistencies, when tasked with complex legal drafting and self-evaluation.

---

[2] C. Engel, Y. Hermstrüwer and A. Kim, 'Human Realignment: An Empirical Study of LLMs as Legal Decision Aids in Moral Dilemmas' Discussion Papers of the Max Planck Institute for Research on Collective Goods (2025); J. Niklaus et al, 'Swiltra-Bench: The Swiss Legal Translation Benchmark' arXiv preprint arXiv:2503.01372 (2025).

[3] Y. Fan et al, 'LEXam: Benchmarking Legal Reasoning on 340 Law Exams' arXiv preprint arXiv:2505.12864 (2025).

Recent works also highlight cross-linguistic challenges. The LLMs' performance is often biased toward English-centric legal systems. Prior studies[15] show that LLMs exhibit cultural biases tied to language use. Among others, Anikina et al., 2025[16] emphasizes the importance of target-language demonstrations, when dealing with non-English or low-resource linguistic contexts, suggesting that LLMs' performance can vary significantly based on the prompt language and the legal tradition involved.

While the aforementioned works focus on university-level exams or bar exams, mostly related to common law systems, our study provides a unique three-tier professional perspective, comparing three distinct career paths in the Italian civil law system, i.e., attorneys, judges, and notaries. Furthermore, by employing a blind "Turing Test" methodology with human experts as evaluators, we move beyond automated scoring, to capture the qualitative nuances of legal "thinking", often missed by current benchmarks. This allows us to identify some specific failures, a boundary where even frontier models may fail to provide legally correct and practically viable solutions. We believe our contribution provides new insights due to the following novelties:

---

[4] E. Martínez, 'Re-Evaluating GPT-4's Bar Exam Performance' *Artificial Intelligence and Law*, 1-24 (2024).
[5] Eric Martínez. 2024. Re-evaluating GPT-4's bar exam performance. Artificial intelligence and law (2024), 1–24.
[6] P.M. Freitas and L.M. Gomes, 'Does ChatGPT Pass the Brazilian Bar Exam?', in *EPIA Conference on Artificial Intelligence* (Cham: Springer, 2023), 131-141.
[7] K. Juvekar, A. Bhattacharya, S. Khadloya and U. Saxena, 'Are LLMs Court-Ready? Evaluating Frontier Models on Indian Legal Reasoning', in N. Aletras et al eds, *Proceedings of the Natural Legal Language Processing Workshop 2025*(Suzhou, China: Association for Computational Linguistics, 2025), 359-369.
[8] Y. Fan et al, 'LEXam' n 3 above.
[9] Narayana et al, 'LEGALBENCH: A Collaboratively Built Benchmark for Measuring Legal Reasoning in Large Language Models', in *Proceedings of the 37th International Conference on Neural Information Processing Systems (NIPS '23)* (Red Hook, NY: Curran Associates Inc, 2023).
[10] Z. Fei et al, 'LawBench: Benchmarking Legal Knowledge of Large Language Models' CoRR abs/2309.16289 (2023).
[11] H. Li et al, 'LexEval: A Comprehensive Chinese Legal Benchmark for Evaluating Large Language Models', in *Proceedings of the 38th International Conference on Neural Information Processing Systems (NIPS '24)* (Red Hook, NY: Curran Associates Inc, 2024), Article 790.
[12] J. Niklaus et al, 'LEXTREME: A Multi-Lingual and Multi-Task Benchmark for the Legal Domain', in H. Bouamor, J. Pino and K. Bali eds, *Findings of the Association for Computational Linguistics: EMNLP 2023* (Singapore: Association for Computational Linguistics, 2023), 3016-3054.
[13] I. Chalkidis et al, 'LexGLUE: A Benchmark Dataset for Legal Language Understanding in English', in S. Muresan, P. Nakov and A. Villavicencio eds, *Proceedings of the 60th Annual Meeting of the Association for Computational Linguistics (Volume 1: Long Papers)* (Dublin, Ireland: Association for Computational Linguistics, 2022), 4310-4330.
[14] R. Pires, R. Malaquias Junior and R. Nogueira, 'Automatic Legal Writing Evaluation of LLMs', in *Proceedings of the Twentieth International Conference on Artificial Intelligence and Law* (2025), 420-424.
[15] Y. Fan et al, 'The Medium Is Not the Message: Deconfounding Text Embeddings via Linear Concept Erasure' arXiv preprint arXiv:2507.01234 (2025); Z. Jin et al, 'Language Model Alignment in Multilingual Trolley Problems' arXiv preprint arXiv:2407.02273 (2024); M.J. Ryan, W. Held and D. Yang, 'Unintended Impacts of LLM Alignment on Global Representation' arXiv preprint arXiv:2402.15018 (2024).
[16] T. Anikina, J. Cegin, J. Simko and S. Ostermann, 'A Rigorous Evaluation of LLM Data Generation Strategies for Low-Resource Languages', in *Proceedings of the 2025 Conference on Empirical Methods in Natural Language Processing* (2025), 8293-8314.

- Multi-Tier Professional Scope: unlike previous work mostly focusing on bar exams, we covered the judicial and notary exams. This comparison elucidates LLMs failures and gaps between doctrinal knowledge and complex, goal-driven legal tasks.
- Expert Blind Evaluation: we employed highly skilled professionals with direct experience in national boards as evaluators, through a "Turing Test" approach, capturing qualitative nuances, often missed by automated metrics.
- Top-Tier Human Benchmarking: for comparing LLMs' and human performances, we relied on the highest-scoring human essays from real national exams as a gold standard, to provide a solid benchmark.
- Long-Form Complex Drafting: we evaluated the ability in writing relatively large, organized, and coherent essays, presenting complex legal contents, rather than short snippets or multiple-choice answers.
- Highly Selective Standards: we applied the same strict criteria used in real exams, where only a small percentage of candidates pass annually, especially as regard judges' and notaries' exams.
- Comprehensive Evaluative Taxonomy: we applied a comprehensive structured grid of criteria, covering language, systemic knowledge, pertinence, coherence and argumentation, to identify specific legal failures in LLMs reasoning.

## III. The Italian Legal Professional Exams

The bar exam evaluates legal reasoning in an adversarial context, including issues identification, persuasive argumentation and the production of procedurally compliant documents. In the assigned task, examinees were required to draft a summons against a company responsible for defective masonry work, seeking both specific
performance (i.e., remediation of the defects) and compensation for damages. The assignment entailed the analysis of complex issues – breach of contractual obligations and liability for damages – and the construction of persuasive arguments, grounded in case law and established legal doctrine, all within a formally valid procedural writ.

The judiciary exam constitutes predominantly a doctrinal reasoning task, within a non-adversarial framework. It requires the impartial analysis of contested legal issues, the systematic exposition of alternative interpretative solutions and the ability to navigate complex doctrinal and jurisprudential debates.

In this study, the assignment was structured in two parts. First, it required an analytical reconstruction of the legal framework governing three unilateral declarations by private parties – acknowledgement of debt, promise to pay, and confession – together with an assessment of their procedural effects. Second, it demanded an examination of their simulation and the legal mechanisms available to revoke the simulation. Unlike contracts, such declarations do not create new legal relations, but confirm pre-existing ones. Therefore, the assignment hinges on a contested doctrinal issue: whether simulated unilateral declarations (e.g., the acknowledgment of a non-existent debt) can produce legal effects and under what conditions such simulations may be revoked. While simulated contracts are generally ineffective under Italian law, the status of simulated non-contractual declarations remains doctrinally unsettled.

Finally, the notary exam reflects a more application-oriented reasoning profile. It requires the drafting of a legal act, addressing the factual situation at issue, motivating the proposed legal solution and discussing the applicable legal institutions and constraints. The act should i) respect all formal and substantial legal requirements, and ii) implement a stable arrangement, enabling all parties to achieve their goals.

In the experimental setting, two assignments were considered, corresponding to an inter vivos and a mortis causa act. The first assignment (inter vivos act) consisted in a complex deed of sale, between a deaf-mute purchaser and a vendor married under the statutory community of property regime. The transaction concerned a building previously donated in fractional shares within a larger compendium. The building was subsequently allocated through a division involving both a cash adjustment and the establishment of an easement, in favour of the property to be sold. The scenario also included the absence of a building permit. The assignment requires evaluating the validity and effectiveness of the prior division, identifying appropriate legal mechanism to protect the purchaser against potential claims by heirs-at-law, and verifying the existence and continuity of the easement. The second assignment (mortis causa act) consisted in drafting a public will, in which the testator intended to appoint an heir, to assign specific assets to both a current spouse and a former spouse, and to separate asset management from ownership. Overall, to implement these objectives required highly contextualized, goal-directed reasoning, combined with doctrinal analysis, strict adherence to all formal requirements applicable to notarial acts, and conformity with the style of notarial practice.

**IV. Methodology and Experimental Setting**

The assigned tests were selected from the most recent examinations available for each profession: the Bar exam administered in December 2024, the Judicial exam administered in January 2024, and the Notary exam administered in May 2023. For each examination, we consulted the official final rankings and obtained the written paper of the candidate who achieved the highest score. The human-authored papers, originally handwritten, were digitized using optical character recognition (OCR). The OCR output was then manually verified and corrected. The resulting documents were rendered as typewritten PDFs and labelled as paper A), B) and C).

Using the same examination prompts, we evaluate four LLMs that, at the time of the experiment, represented the state of the art. Each model is assigned a one-letter identifier: A) Claude 4 Opus, B) GPT-5, D) DeepSeek R1, and E) Gemini 2.5 Pro. At the time of testing, Gemini 2.5 Pro had just been released and its built-in safeguards prevented the generation of legal texts tout court. We overcame this limitation by submitting the prompts through the model's API. API calls did not play with model options (e.g., generation temperature), thus soliciting default behaviour. For all models, we enabled the newly introduced thinking mode and allowed access to internet search. By inspecting the list of reported sources, we verified that none of the generated outputs reproduced solutions retrieved from specialized online repositories. Finally, references to case law were removed from all documents to ensure uniformity and to make the LLM-generated outputs indistinguishable from the human-authored papers, which did not contain such references (as it is usually the case in these exams).

Tests were submitted using minimal prompts, sharing a simple and uniform structure. Each prompt consisted of four components: (i) a declaration of the model's role ("You are a candidate for . . . "), (ii) a task description ("Generate the document in response to the question below, taking into account all the provided information"), (iii) a verbatim reproduction of the exam question and (iv) a set of specific instructions. The first part of the instruction section was common to all three examinations and imposed the following constraints: the generated text had to include explicit references to relevant legislation; avoid unnecessary repetitions; and it could be organized into sections, but not structured using multi-level bullet points. An approximate length requirement of 1500 words was added for the Bar and Judicial exams, corresponding to the usual length of exam papers by humans. Specific concise requirements were added for each type of paper. All prompts and materials can be found in our repository.[17] As an example, consider the following prompt, used for the Judicial Exam.

<persona>
You are a candidate in a Judicial exam in Italy.
</persona>
<task>
Generate the discussione required by the following prompt.
</task>
<prompt>
After discussing the acknowledgement of debt, promise to pay,

[17] https://github.com/SSIGPRO/LLMs-in-legal-professions

and confession and their procedural consequences,
the candidate should focus on the possibility of simulating
such acts and the possibility of a knowingly untrue
confession.
</prompt>
<instructions>
The generated text must be approximately 1,500 words.
The generated text must contain explicit references to
relevant statutes.
The generated text must avoid repetitions.
The generated text may contain references to case law.
For every judgment cited, verify that it exists and is
pertinent.
The generated text may contain sections but must not be
structured with multi-level bullet points.
The generated text must meet the following criteria:
1) It must demonstrate correct Italian in terms of
terminology, syntax, and grammar, and show adequate
mastery of legal terminology and sufficient clarity of
exposition—all of which are indispensable requirements
for the correct drafting of judicial measures;
2) It must present a pertinent, coherent, and exhaustive
treatment of the assigned topic, demonstrating the
candidate's sufficient knowledge of the institutions
directly referred to and the fundamental principles
of the subject, as well as an adequate general legal
culture;
3) It must reveal the candidate's ability to analyze the
underlying issues of the topics proposed by the prompt,
demonstrating a critical awareness of normative sources
and explaining the outcome of the legal reasoning; such
reasoning, even if not shared [by the examiner], shall
not be a determining factor in the evaluation as long as
it is logically argued and consistent with the
institutions and principles of the field.
</instructions>

Texts generated by the four LLMs in response to their respective prompts were formatted identically to the human-authored papers, making them visually indistinguishable. Evaluators were selected among those who had previously served on the national board committees, in past examination sessions. They were provided with five uniformly formatted papers, with no indication of whether a given document had been produced by a human candidate or by a specific language model.

## V. Evaluation Criteria

All LLMs-generated and human exam-papers were independently evaluated by the three domain experts. They were asked to assess five exam papers: one drafted by a human candidate (who successfully passed the exam) and 4 generated by LLMs, namely Claude v4 Opus by Anthropic, ChatGPT v5 by OpenAI, DeepSeek R1 by Deep Seek, and Gemini v2.5 pro by Alphabet.

To ensure comparability across evaluators, we designed structured criteria, based on a set of dimensions and a common scoring scale. In defining the criteria we relied on official guidelines issued by the Italian Ministry of Justice, for each exam. Whenever a numerical score was required, domain experts used a four-point scale from 0 to 3, respectively meaning *insufficient* (below the minimum standard for the exam), *sufficient* (meeting the minimum standard without distinctive strengths), *good* (clearly above the minimum standard though not exceptional), and *excellent*. Following the official guidelines, in addition to numerical scores, experts were requested to apply a set of binary checks, meant to assess compliance with essential formal requirements of the legal documents to be drafted, in the bar and notary exams.

**1.The Bar Exam**

The Bar exam concerned drafting a pleading for a legal proceeding. The evaluation form addressed linguistic quality and substantive legal content (see Table 1 for the detailed results) as well as the formal requirements of legal summons. Linguistic quality was measured along two dimensions.

The first, linguistic correctness, covered: (i) grammar, syntax, orthography; and (ii) the appropriate use of legal terminology. The second, clarity and rigour, covered: (i) coherence, completeness and non-redundancy (ii) methodological rigour in legal argument (iii) relevance to the assigned topic.

Substantive legal analysis was evaluated along two main dimensions. The first concerned the ability to address the legal issues at stake, including: (i) knowledge of theoretical legal foundations (ii) he applicable normative framework, (iii) references to doctrinal and case-law orientations, (iv) legal reasoning. The second dimension concerned argumentation and persuasion: (i) the ability to support legal conclusions and (ii) the mastery of argumentative and persuasive techniques.

The formal requirements concerned mandatory elements of a summons act: (i) curia adita (the court being addressed), (ii) the parties, (iii) the factual and legal grounds, (iv) the summoning of the defendant, (v) the conclusions, (vi) date and signature, (vii) service notification, and (viii) the power of attorney.

**2. The Judicial Exam**

As noted, the judicial exam involves the examination of a broad legal issue in the context of legal norms, case law and doctrinal debate.

The assessment form specified criteria concerning: linguistic quality; knowledge of legal institutes and principles, together with general legal culture; and the capacity to analyse legal issues. Linguistic quality was assessed using binary labels, while the remaining aspects were scored on a 0–3 scale (see Table 2 for the detailed results). Linguistic quality included :(i) terminological, syntactic, and grammatical correctness; (ii) command of legal terminology; and (iii) clarity of exposition.

The requirements on knowledge of legal institutes and principles covered: (i) knowledge of relevant legal institutes, including awareness of competing theories, interpretations, and applications; (ii) the ability to contextualise and organise such knowledge, connecting specific hypotheses to general legal categories; (iii) pertinence, coherence and exhaustiveness.

The requirements on case solution included: (i) critical awareness of normative sources; (ii) logical argumentation, (iii) coherence with relevant legal principles and institutes.

**3. The Notary Exam**

The notary exam has a more practical oriented focus and consists in drafting a notary deed (typically a contract or will), accompanied by a discussion of the relevant legal issues. For both mortis causa and inter vivos deeds, the assignment aims at assessing the ability to translate a complex factual and legal scenario into a rigorously reasoned and formally valid notarial deed.

The corresponding form included i) a set of binary checks aimed at verifying the satisfaction of certain minimal necessary requirements, and ii) a set of qualitative requirements to be numerically graded (only the latter are included in Tables 3 and 4.

The binary checks concerned the following:(i) misreading or misinterpretation of the case; (ii) incompleteness of the deed; (iii) contradictory legal arrangements (iv) inconsistency between legal arrangements and their justification; (iv) omission or insufficient treatment of legally relevant institutes; (vi) errors of law within the deed or (vii) the theoretical discussion; (viii) significant linguistic inaccuracies (e.g., grammatical, syntactic, or orthographic errors); (ix) violation of formal requirements governing notarial deeds; (x) inadequacy of the deed in relation to the parties' interests and objectives; (xi) deficiencies in completeness, logical coherence, order, clarity or legal correctness.

The qualitative evaluation included: (i) the identification of legally relevant institutes and (ii) of the parties' interests; (iii) the adequacy of the selected legal institute; (iv) the justification of that selection; (v) the applicability of the proposed legal solution to the case; (vi) the quality of the deed's drafting; and (vii) the discussion of relevant institutes.

## VI. Results and Discussion

This section presents and discusses the obtained results, following the criteria outlined in Section V.

### 1. The Bar exam

Results reveal substantial variability, as the overall performance – measured as the cumulative score assigned by domain experts across all the evaluated dimensions – ranges from a minimum of 26 to 79 out of 100. Detailed results are reported in Table 1.

Two models significantly surpass the human candidate (who scored 62): Gemini 2.5 Pro achieved the highest overall score (79), followed by ChatGPT-5 (65). According to the scoring grid in the experimental data, their performance highlights several markers of advanced adversarial reasoning, in particular under the argumentation and persuasion dimensions. This suggests that such models do not merely synthesize the applicable law; they can also build sound and persuasive arguments supporting legal claims. Notably, Gemini 2.5 Pro outperforms the human benchmark across all evaluation steps and macro-criteria.

Regarding linguistic quality, clarity, and doctrinal robustness (Steps 1–2), evaluators consistently reported high levels of grammatical accuracy, terminological precision, and structural organisation. The generated pleading exhibits a clearly articulated argumentative structure, a precise reconstruction of the applicable normative framework, as well as internally consistent legal reasoning. In several dimensions –particularly methodological rigour and normative completeness – Gemini is rated above the typical standard of top-tier human candidates. In an adversarial context, this may enable the model to anticipate and preemptively address potential counterarguments, providing a litigant with an accurate targeted reconstruction of the applicable legal framework. In line with these results, as regard the verification of formal requirements (Step 3), Gemini correctly identified and implemented 7 out of 8 mandatory elements. The only omission is the "service notification (relata di notifica)" – a specific certification of transmission to the defendant. This high degree of formal compliance is particularly noteworthy given that models are tested out-of-the-box (with no legal fine-tunig or RAG) and did not receive any explicit prompting regarding the compulsory components of Italian pleadings. Conversely, DeepSeek R1 exhibits critical deficiencies across all dimensions, showing significant underperformance.

Overall, results show pronounced differences in performance across models, both in terms of cumulative scores and in the distribution of strengths and weaknesses across linguistic, substantive, and formal evaluation criteria. More generally, while human candidates maintain a slight edge in "doctrinal nuance", the best performing models' capacity for structured legal reasoning and formal reconstruction indicates that they can be powerful assets for drafting persuasive legal documents in the Italian jurisdiction.

### 2. The Judicial exam

Results of the judicial exam provide a clear benchmark for systemic doctrinal analysis, i.e., the ability to contextualize specific legal institutes within the broader legal system and to navigate complex theoretical debates with neutrality. Detailed results are reported in Table 2.

Gemini 2.5 Pro emerged as the outlier, not only achieving the highest score (21/24), but also demonstrating a capacity for legal analysis above the top-tier human benchmark (18/24). Evaluators noted several key markers of systemic legal knowledge: accurate command of the doctrinal landscape, precise handling of competing legal interpretations, and consistently coherent exposition.

Indeed, Gemini matched the top human performance with respect to knowledge of relevant legal institute and the ability to relate such knowledge to more general legal concepts. This suggests that LLMs can provide a more comprehensive overview of doctrinal and judicial opinions than a human candidate under exam pressure. The model also equalled the human in awareness of legal sources – including correct and pertinent citations of legislation – as well as in argument consistency and coherence with established legal institutes and principles. Finally, Gemini outperformed the human candidate in applying legal knowledge to the specific essay topic, in terms of pertinence, coherence and exhaustiveness.

The remaining models' performance was substantially worse, failing to move beyond basic linguistic fluency. ChatGPT-5 obtained a 10 score, reflecting deficiencies in pertinence and exhaustiveness, as well as weaknesses in the discussion of legal sources and in the logical stringency of the argumentation. Claude 4 Opus and DeepSeek R1 failed even more significantly in the substantive steps.

With the exception of DeepSeek R1, all models satisfied linguistic quality requirements (Step 1). Claude 4 Opus, ChatGPT-5, and Gemini 2.5 Pro demonstrated adequate command of syntax, grammar and legal terminology, as well as overall clearity of exposition.

Overall, the results of the Judicial exam revealed substantial variability across models and a clear separation between the highest-performing one and the remaining LLMs, as well as between LLMs performance and the human benchmark.

**3. The Notary exam**

The notary exam proved to be the most challenging setting for all tested LLMs. Unlike the Bar and Judicial exams – where some models excelled and others delivered satisfactory results– all LLMs show a "performance ceiling". While models showed proficiency in adversarial argumentation (Bar) and doctrinal synthesis (Judicial), they consistently failed to meet the standards of goal-directed legal planning required for notarial practice. All LLMs fell significantly below the human benchmark across both the inter vivos and mortis causa assignments.

In both cases, evaluation followed a two-step procedure. The first one assesses compliance with a set of mandatory requirements. The second step evaluates the quality of the legal analysis and the proposed solution. Although the inter vivos and mortis causa assignments differ in their doctrinal foundations and formal requirements, the overall performance across models is consistent: the human candidate clearly outperforms all LLMs-generated papers. Across both assignments, all models demonstrated adequate command of the Italian language and basic legal terminology. However, they failed to pass both evaluation steps. Detailed results are shown in Table 3 and 4.

With regard to the inter vivos assignment, only the human candidate satisfied all the mandatory requirements (Step 1). Unlike the Bar exam, Gemini 2.5 Pro failed 9 out of 12 requirements. GPT 2.5 follows with 8.

In the evaluation of legal analysis, the human essay obtained 14 out of 21 points from each evaluator, for a total of 42. The best-performing model, Gemini 2.5 Pro, achieved only 5 points, according to the two most lenient evaluators (total: 10). Among LLMs, Gemini 2.5 Pro was relatively more accurate in identifying relevant legal institutions and justifying its proposed approach, whereas ChatGPT-5 produced the most internally coherent overall solution. Claude 4 Opus performed comparably to Gemini 2.5 Pro in the evaluation of necessary requirements (Step 1), but scored lower than Gemini 2.5 Pro and GPT-5 in legal quality, showing only minimal ability to identify the parties' interests and relevant institutions, and failing to adequately develop the remaining aspects of the assignment. DeepSeek R1 obtained a 0 score, failing to meet even the minimal expectations across both evaluation Steps. These results underscore that LLMs struggle to move from "discussing" law to "building" legally sound, risk-free instruments.

In the mortis causa acts, only the human candidate performed at a sufficient level (meeting all necessary requirements, and scoring 14 points on in legal quality from each evaluator, for an aggregate score of 42), though LLMs performed better than in the inter vivos. With regard to necessary requirements GPT 5 was the best model, failing only two of them. It also outperformed the other models in legal quality, with a total score of 20 (compared to 8 for the two next best-performing models). Two evaluators considered the quality of the GPT 5 paper similar to the human-authored paper, with regard to the adopted solutions and their justification, the identification of the relevant legal institutions, the legal qualification of the case, and the overall drafting of the will. GPT 5 is followed by Gemini 2.5 Pro, which exhibited a broadly similar performance pattern across the two notarial assignments. Claude 4 Opus ranked next, achieving a slightly better result than the inter vivos act, narrowing the gap with Gemini 2.5 Pro. DeepSeek R1 once again proved consistently insufficient across all evaluated dimensions.

Overall, results show that none of the evaluated LLMs succeeded in producing notarial deeds, simultaneously satisfying mandatory formal requirements and an adequate level of substantive legal quality.

**VII. Comparison across legal tasks and performances**

Results highlight a strong dependence of LLMs performance on the nature of legal tasks, required by each exam. Across the three examination, their performance varies systematically as tasks shift from adversarial argumentation to doctrinal exposition and to highly goal-oriented legal drafting.

In the Bar and Judicial exams, high-performing models demonstrated substantial competence in tasks centred on the organisation, articulation, and application of existing legal knowledge. These settings primarily require the identification of legally relevant issues, the reconstruction of normative frameworks and the production of coherent and well-structured legal arguments. Under such conditions, frontier LLMs are able to approximate – and in some cases exceed – human benchmark performance.

By contrast, the notary exam exposes a limitation in current LLMs capabilities. Unlike the other exams, notarial drafting requires the coordinated selection of legal instruments, the anticipation of downstream legal effects, the design of a suitable legal arrangement and compliance with many formal and procedural constraints. The LLMs failure in this setting indicates that success in argumentative or doctrinal legal tasks does not readily transfer to contexts requiring long-horizon legal planning and risk-sensitive instrument design.

Taken together, these findings suggest that LLM performance in the legal domain cannot be assessed in the abstract, but must be evaluated in relation to specific tasks, including the degree of knowledge accessibility, the orientation of legal reasoning required, and the density of formal and instrumental constraints.

Beyond task-specific effects, results reveal substantial variability in performance across the evaluated LLMs. This variability is systematic, affecting all examined dimensions – from linguistic adequacy to substantive legal analysis and formal correctness – and leading to sharply divergent outcomes under identical evaluation conditions.

At one end of the spectrum, Gemini 2.5 Pro consistently achieved the highest scores across all exams, in some cases exceeding the human benchmark. Its performance was characterised by stability across evaluation criteria, with no single dimension emerging as a dominant point of failure. At the other end, DeepSeek R1 consistently failed to meet minimum thresholds, particularly in tasks requiring engagement with legal institutes, doctrinal sources, and structured legal reasoning. The other models, such as ChatGPT-5 and Claude 4 Opus, displayed uneven performance profiles, achieving adequacy in some dimensions, while exhibiting marked weaknesses in others.

Crucially, these differences cannot be reduced to surface-level fluency. While most models satisfied basic linguistic requirements in the Bar and Judicial exams, substantive legal competence and formal reliability varied widely. In particular, the gap between top-performing and lower-performing systems widens as tasks demand deeper legal analysis, stricter adherence to formal constraints, or coordinated planning across multiple legal dimensions.

This heterogeneity has important implications for the reliability of LLMs in legal contexts. The presence of a small number of high-performing systems does not imply that LLMs as a class can be treated as uniformly capable. Rather, performance appears to be highly model-dependent, with sharp discontinuities even among systems that are comparable in size or general-purpose capabilities.

Accordingly, any assessment of LLMs suitability for legal tasks must consider not only average performance, but also variance, failure modes and sensitivity to task structure. Treating LLM outputs as interchangeable or broadly representative of "AI performance" in the legal domain risks obscuring critical differences that bear directly on deployment, oversight and risk management.

**VIII. Failure Patterns in Notary Exam**

The notary exam constitutes an extreme – and therefore particularly informative – test case for analysing LLMs limitations in the legal domain. Unlike the Bar and Judicial exams, notarial drafting combines strict formal constraints with goal-oriented legal planning and direct sensitivity to the parties' interests. For this reason, the underperformance of all evaluated LLMs in the Notary exam warrants dedicated analysis beyond aggregate performance scores.

Based on expert evaluations, and as a pilot analysis, we identified the most recurrent errors, to which refer to as legal failures.

The classification of such legal failures is developed through a purpose-built taxonomy, jointly considering: (i) the requirements of notarial assignments; and (ii) the soundness of the underlying legal reasoning, including consistency between the proposed solutions, their justification, and theoretical discussion. On this basis, we distinguish five main categories:

(1) Legal-source failure, consisting of references to inapplicable, incorrect, or fictitious legal rules, as well as arbitrary or forced interpretations of applicable norms;

(2) Reasoning failure, understood as internal inconsistency in the system's output with respect to the exam prompt, external data, or established notarial practice;

(3) Pertinence-failure, consisting in the inability to take a position on, or the elusion of, a central legal question raised by the assignment;

(4) Lexical failures, i.e. the improper or imprecise use of technical legal terminology;

(5) Formal failures, namely structural deficiencies of the deed, or stylistic and drafting defects affecting its notarial correctness.

All LLMs exhibits some legal failures in both assignments. In the inter vivos act, even Gemini 2.5 Pro – the best performing model – exhibited some deficiencies across all categories, with a predominance of formal and reasoning failures. In the mortis causa assignment, ChatGPT-5 – the best-performing model – incurred in interpretative inaccuracies (reasoning failures) and residual formal defects.

Despite differences across assignments, none of the evaluated LLMs succeeded in satisfying all mandatory formal requirements and in reaching a sufficient legal quality, in each of the two notarial tasks. This confirms that, in their current out-of-the-box configuration, models are unable to produce notarial acts that are simultaneously legally sound, formally valid, and adequately aligned with the parties' interests, at least with respect to the complex scenarios typical of highly selective notarial examinations.

At the same time, the analysis also reveals residual strengths. Across both assignments, most LLMs were able to identify at least some relevant legal institutes, recognise the basic interests of the parties or the testator, and maintain acceptable linguistic clarity and terminological adequacy. These suggests that current LLMs may be useful in the notary domain as support tools in preliminary drafting stages or in educational contexts, provided that their outputs remain subject to strict human supervision.

Finally, the contrast between inter vivos and mortis causa performance offers further insight into task sensitivity. In the mortis causa assignment, LLMs respected a greater number of essential requirements, committed fewer errors of law, and produced more coherent and internally consistent reasoning. This difference plausibly reflects the lower structural complexity of mortis causa scenarios, typically involving a single party's intentions within a more stable legal framework. By contrast, inter vivos transactions require the coordination of multiple parties' interests, prior legal acts, and conditional safeguards designed to operate across time-features that appear to exceed the current planning and contextual reasoning capacities of LLMs.

## IX. Some speculations on the results of the experiment

The different scores obtained by the LLMs in the three exams –in particular the difference between the Bar and Judicial exams in comparison to the Notary exams – invite some speculations on the underlying grounds.

First, it seems that the legal performance of an LLMs may depend on the amount of relevant knowledge that is easily accessible and thus can be provided to the system either at the pre-training phase or through online searches. In fact, the topics addressed in the Bar and Judicial exams concern doctrinal issues on which vast legal knowledge is available on easily accessible textbooks and cases. On the contrary, private contracts written by Italian notaries are generally not publicly available, so that the LLMs could access only a limited number of relevant examples. The high performance of LLMs in drafting goal-driven legal arguments in the context of summons could also be facilitated by the fact that such documents are largely available online, for the kind of cases under consideration (traffic accidents).

A second aspect that may contribute to explain the different performance relates to the fact that current LLMs already excel at selecting, synthesising and organising relevant knowledge, but are less proficient in goal-driven problem solving.

Indeed, the focus of the Bar – and even more so of the Judicial – exams was on presenting a comprehensive picture of the legal landscape surrounding the assigned topic. By contrast, the Notary exam focused on the preparation of highly contextualised legal acts. Their drafting (in particular, the inter-vivos contract) required a planning exercise meant to identify the parties' goal and to select solutions best suited to achieve such goals. Moreover, the doctrinal considerations required in the notary exam were expected to be highly pertinent to the specific issues raised by the contract.

The LLMs limitation in goal-driven and contextualised understanding may also help explain why LLMs were misled by certain tricks and snares embedded in the Notary exam (though rarely encountered in real-world practice). For instance, the inter-vivos assignment involved a property transfer to an individual who was deaf-mute but able to read and write. Under notarial law, such a person must read and sign the deed, and the presence of a sign-language interpreter is mandatory under penalty of nullity (whereas an interpreter would not be required if the individual were deaf but not mute). All models, including those that correctly identified legal sources, incorrectly stated that an interpreter was not needed, and failed to mention one in the deed. By falling into such a trap, LLMs exhibited a considerable degree of gullibility. This, in turn, highlights the risk that LLMs may be deceived, misled, or manipulated when deployed in legal tasks, suggesting the need for particular caution in the adoption of agentic AI within the legal domain.

## X. Limitations

Results should be interpreted in light of limitations inherent in the experimental design. These limitations do not undermine the validity of our findings, but rather circumscribe their scope and inform the extent to which our conclusions can be generalised.

First, the experimental conditions under which LLMs are evaluated differ from those faced by human candidates in official examinations. In particular, LLMs were deployed with access to Internet, whereas human candidates – under exam conditions – are typically restricted to a very limited set of legal materials, usually just the Italian civil code. This asymmetry may influence performance, especially in tasks requiring extensive doctrinal or jurisprudential recall. At the same time, it should be noted that the LLMs were used out of the box, without task-specific legal fine-tuning or dedicated retrieval-augmented generation (RAG), while human candidates generally prepare for these exams over extended periods.

Second, the evaluation relies on a limited number of exam papers and on a small pool of expert evaluators, reflecting the structure of the real examinations but constraining statistical generalisation. While the use of official assessment grids and experienced examiners enhances ecological validity, the sample size does not support fine-grained quantitative claims about performance distributions beyond the cases considered.

Finally, the analysis is situated within the Italian legal systems and institutional context. Differences in legal traditions, procedural frameworks, professional training, and exam practice may affect the extent to which these results transfer to other jurisdictions or to transnational legal settings.

Taken together, these limitations suggest that the findings should be read as a structured exploration of LLM capabilities under exam-like conditions, rather than as a comprehensive assessment of their potential role in legal practice.

In this regard, however we must underline some limitations of our experiment:

- The used LLMs reflect the state of this rapidly evolving technology at the time of our experiment (September 2025);
- They were deployed out-of-the-box, without any special legal fine-tuning or focused RAG.
- They were provided with a single simple prompt for each of the exams under scrutiny (bar, judges, inter-vivos and mortis-causa).
- The provided prompts were very simple, just asking to answer the assignment taking the role of a candidate.
- The evaluations were entrusted only to three examiners (between which, however, a relatively strong agreement could be found).

## XI. Conclusion and Future Works

Our experiment shows that out-of-the-box LLMs exhibit considerable degree of legal competence, while still facing significant limitations.

Results from the Bar and Judicial examinations show that LLMs excel in general legal knowledge, even when complex doctrinal issues are involved, areas in which only humans with extensive, in-depth legal knowledge perform proficiently. Models are able to easily process the vast body of knowledge acquired during pre-training, expand it with information available online, and apply the resulting knowledge to the case at hand. By contrast, results in the Notary examination, reveal that LLMs may struggle with highly contextualised legal tasks, particularly in domains where limited background knowledge is available and where goal-directed reasoning is required.

These conclusions emerge from the blind assessment of examination papers by LLMs and human candidates, conducted by professional examiners. Despite these limitations, we believe that our experiment provides meaningful insights on the capacities and limitations of LLMs in performing legal tasks, as well as into the opportunities and risks associated with their use in legal practice. We should keep in mind that performance in a professional exam does not entail the ability to exercise the profession at stake. For instance a judge needs to engage in many cognitive tasks that go beyond the ability to develop a coherent doctrinal analysis (as required by the judges' exam). The judge needs to be able to interact with the parties and their laywers, to assess the relevant facts, to understand social interactions and the interest at stake, to interpret the law according to the purposes and values of the legal system, etc. For human candidates the ability to write good doctrinal essays can be taken as a proxy for the possession of (or the capacity to obtain) the abilities required to be a good judge. This is clearly not the case for an LLM. On the other end, the ability of the LLMs to provide good reviews of case law and doctrine (as in the judges' exam), and to organize legal material in rethorical-dialectical perspective (as in the lawyers' exam), indicate possible directions for their use for supporting legal professionals.

We aim to extend our research along several directions, including: evaluating new releases of a broader set of LLMs, comparing out-of-the-box models with LLM-based systems specifically tailored for legal applications, assessing LLMs' performance using a more fine-grained set of evaluation criteria, and combining human assessments with the use of LLMs-as-a-judge, across multiple benchmarks.

Furthermore, we plan to investigate the distinguishability of human-authored legal documents from those generated by LLMs, focusing on the perspective of general legal practitioners rather than expert evaluators. To this end, we will reuse the existing corpus and design a Turing-like experiment, in which legal professionals will be asked to identify the human-authored text among AI-generated documents. We anticipate heterogeneity in the level of attention and effort that participants will devote to the task. This variability will enable a stratified analysis of whether, and to what extent, human authorship can be detected under varying levels of evaluative engagement. Such an investigation will provide a meaningful assessment in an environment increasingly dominated by AI-generated content, where identifying, appreciating and valuing human contributions is likely to become both more challenging and more significant.

| | Claude 4 Opus | | | | GPT 5 | | | | Human | | | | DeepSeek R1 | | | | Gemini 2.5 Pro | | | |
|---|---|---|---|---|---|---|---|---|---|---|---|---|---|---|---|---|---|---|---|---|
| | #1 | #2 | #3 | tot | #1 | #2 | #3 | tot | #1 | #2 | #3 | tot | #1 | #2 | #3 | tot | #1 | #2 | #3 | tot |
| **Correctness** | | | | | | | | | | | | | | | | | | | | |
| Grammatical form, syntax, spelling | 3 | 3 | 2 | **8** | 3 | 3 | 3 | **9** | 3 | 3 | 2 | **8** | 2 | 3 | 1 | **6** | 3 | 3 | 3 | **9** |
| Legal terminology | 2 | 2 | 2 | **6** | 2 | 3 | 3 | **8** | 2 | 3 | 2 | **7** | 1 | 3 | 2 | **6** | 2 | 3 | 3 | **8** |
| **Clarity and rigour** | | | | | | | | | | | | | | | | | | | | |
| Coherence, completeness, non redundancy (clarity in exposition) | 1 | 2 | 0 | **3** | 1 | 2 | 3 | **6** | 2 | 3 | 1 | **6** | 0 | 2 | 0 | **2** | 2 | 3 | 3 | **8** |
| Methodological rigour of exposition and legal arguments | 1 | 2 | 0 | **3** | 1 | 2 | 3 | **6** | 2 | 3 | 2 | **7** | 0 | 1 | 0 | **1** | 2 | 3 | 2 | **7** |
| Relevance of the paper to the assigned topic | 1 | 2 | 1 | **4** | 1 | 2 | 3 | **6** | 2 | 3 | 2 | **7** | 0 | 1 | 1 | **2** | 2 | 3 | 3 | **8** |
| **Capacity to resolve specific legal issues** | | | | | | | | | | | | | | | | | | | | |
| Knowledge of the theoretical foundations | 1 | 2 | 1 | **4** | 2 | 2 | 3 | **7** | 2 | 2 | 2 | **6** | 0 | 1 | 2 | **3** | 2 | 3 | 2 | **7** |
| Identification of the applicable legal framework | 0 | 1 | 1 | **2** | 0 | 1 | 2 | **3** | 1 | 2 | 2 | **5** | 0 | 1 | 0 | **1** | 2 | 3 | 3 | **8** |
| References to relevant doctrine and case-law | 0 | 0 | 1 | **1** | 1 | 1 | 3 | **5** | 0 | 1 | 1 | **2** | 0 | 0 | 0 | **0** | 0 | 1 | 2 | **3** |
| Presentation of the legal reasoning and conclusions | 1 | 1 | 1 | **3** | 1 | 2 | 2 | **5** | 2 | 3 | 2 | **7** | 0 | 1 | 0 | **1** | 2 | 3 | 3 | **8** |
| **Demonstration of argumentative and persuasive capacity** | | | | | | | | | | | | | | | | | | | | |
| Ability to adequately support the conclusions | 1 | 1 | 0 | **2** | 1 | 2 | 2 | **5** | 1 | 2 | 1 | **4** | 0 | 1 | 1 | **2** | 2 | 2 | 3 | **7** |
| Mastery of argumentative and persuasive techniques | 2 | 1 | 2 | **5** | 1 | 2 | 2 | **5** | 1 | 1 | 1 | **3** | 0 | 1 | 1 | **2** | 2 | 2 | 2 | **6** |
| **Total** | **13** | **17** | **11** | **41** | **14** | **22** | **29** | **65** | **18** | **26** | **18** | **62** | **3** | **15** | **8** | **26** | **21** | **29** | **29** | **79** |

Table 1: Grid for the bar-exam papers exam as filled by expert #1, #2, and #3.

| | Claude 4 Opus | GPT 5 | Human | DeepSeek R1 | Gemini 2.5 Pro |
|---|---|---|---|---|---|
| | #1,2,3 | #1,2,3 | #1,2,3 | #1,2,3 | #1,2,3 |
| **Requirements on linguistic quality** | | | | | |
| Correct linguistic form (terminology, syntax, and grammar) | Yes | Yes | Yes | Yes | Yes |
| Adequate command of legal terminology | Yes | Yes | Yes | Yes | Yes |
| Sufficient clarity of exposition | Yes | Yes | Yes | No | Yes |
| **Requirements on knowledge of institutes and principles** | | | | | |
| Knowledge of sources and theories | 1 | 2 | 3 | 0 | 3 |
| Systemic understanding | 1 | 2 | 3 | 0 | 3 |
| In the discussion of the case: | | | | | |
| Pertinence | 1 | 1 | 2 | 0 | 3 |
| Coherence | 1 | 2 | 2 | 0 | 3 |
| Exhaustiveness | 0 | 0 | 2 | 0 | 3 |
| **Requirements on the case solution** | | | | | |
| Critical awareness of normative sources | 0 | 1 | 2 | 0 | 2 |
| Presentation of the outcome of the legal reasoning: | | | | | |
| Logical argumentation | 0 | 1 | 2 | 0 | 2 |
| Coherence with the relevant legal institutes and principles | 0 | 1 | 2 | 0 | 2 |
| **Total** | **12** | **30** | **54** | **0** | **63** |

Table 2: Grid for the judges' exam (agreed upon marks are weighted by the number of concurring evaluators).

| | Claude 4 Opus | | | GPT 5 | | | Human | | | DeepSeek R1 | | | Gemini 2.5 Pro | | |
|---|---|---|---|---|---|---|---|---|---|---|---|---|---|---|---|
| | #1 | #2,3 | tot | #1 | #2,3 | tot | #1 | #2,3 | tot | #1 | #2,3 | tot | #1 | #2,3 | tot |
| **Failure reasons** | | | | | | | | | | | | | | | |
| Misunderstood assignment | No | No | | Yes | Yes | | No | No | | Yes | Yes | | Yes | No | |
| Incompleteness | Yes | Yes | | Yes | Yes | | No | No | | Yes | Yes | | Yes | Yes | |
| Unsuitable solutions | Yes | Yes | | Yes | No | | No | No | | Yes | Yes | | Yes | Yes | |
| Inconsistent solutions | Yes | Yes | | Yes | No | | No | No | | Yes | Yes | | Yes | Yes | |
| Contradiction between solutions and reasons | No | No | | Yes | No | | No | No | | Yes | Yes | | Yes | No | |
| Insufficient doctrinal analysis | Yes | Yes | | Yes | Yes | | No | No | | Yes | Yes | | Yes | Yes | |
| Legal mistakes in the act | Yes | Yes | | Yes | Yes | | No | No | | Yes | Yes | | Yes | Yes | |
| Legal mistakes in the theoretical analysis | Yes | Yes | | Yes | Yes | | No | No | | Yes | Yes | | Yes | Yes | |
| Language mistakes | No | No | | Yes | No | | No | No | | No | No | | No | No | |
| Formal mistakes | Yes | Yes | | Yes | Yes | | No | No | | Yes | Yes | | Yes | Yes | |
| Unsuitability to the parties' interests | Yes | Yes | | Yes | Yes | | No | No | | Yes | Yes | | Yes | Yes | |
| Insufficient motivation | Yes | Yes | | Yes | Yes | | No | No | | Yes | Yes | | Yes | Yes | |
| **Assessment criteria** | | | | | | | | | | | | | | | |
| Identification of legal institutes | 0 | 1 | **2** | 0 | 1 | **2** | 2 | 2 | **6** | 0 | 0 | **0** | 0 | 1 | **2** |
| Identification of the parties' interests | 0 | 1 | **2** | 0 | 1 | **2** | 2 | 2 | **6** | 0 | 0 | **0** | 0 | 1 | **2** |
| Adequacy of the selected legal institutes | 0 | 0 | **0** | 0 | 0 | **0** | 2 | 2 | **6** | 0 | 0 | **0** | 0 | 1 | **2** |
| Justification of that selection | 0 | 0 | **0** | 0 | 1 | **2** | 2 | 2 | **6** | 0 | 0 | **0** | 0 | 1 | **2** |
| Suitability of the solution | 0 | 0 | **0** | 0 | 1 | **2** | 2 | 2 | **6** | 0 | 0 | **0** | 0 | 1 | **2** |
| Quality of the deed's drafting | 0 | 0 | **0** | 0 | 0 | **0** | 2 | 2 | **6** | 0 | 0 | **0** | 0 | 0 | **0** |
| Discussion of legal institutes | 0 | 0 | **0** | 0 | 0 | **0** | 2 | 2 | **6** | 0 | 0 | **0** | 0 | 0 | **0** |
| **Total** | **0** | **2** | **4** | **0** | **8** | **8** | **14** | **28** | **42** | **0** | **0** | **0** | **0** | **10** | **10** |

Table 3: Grid for the inter vivos notary exam (agreed-upon marks are weighted by number of concurring evaluators).

| | Claude 4 Opus | | | GPT 5 | | | Human | | | DeepSeek R1 | | | Gemini 2.5 Pro | | |
|---|---|---|---|---|---|---|---|---|---|---|---|---|---|---|---|
| | #1 | #2,3 | tot | #1 | #2,3 | tot | #1 | #2,3 | tot | #1 | #2,3 | tot | #1 | #2,3 | tot |
| **Legal failures** | | | | | | | | | | | | | | | |
| Misunderstood assignment | Yes | No | | Yes | No | | No | No | | Yes | Yes | | Yes | Yes | |
| Incompleteness | Yes | No | | Yes | No | | No | No | | Yes | Yes | | Yes | Yes | |
| Unsuitable solutions | Yes | Yes | | Yes | No | | No | No | | Yes | Yes | | Yes | No | |
| Inconsistent solutions | Yes | No | | Yes | No | | No | No | | Yes | Yes | | Yes | No | |
| Contradiction between solutions and reasons | Yes | Yes | | Yes | No | | No | No | | Yes | Yes | | Yes | No | |
| Insufficient doctrinal analysis | Yes | Yes | | Yes | Yes | | No | No | | Yes | Yes | | Yes | Yes | |
| Legal mistakes in the act | Yes | Yes | | Yes | No | | No | No | | Yes | Yes | | Yes | Yes | |
| Legal mistakes in theoretical analysis | Yes | No | | Yes | No | | No | No | | Yes | Yes | | Yes | Yes | |
| Language mistakes | No | No | | No | No | | No | No | | No | No | | No | No | |
| Formal mistakes | Yes | Yes | | Yes | Yes | | No | No | | Yes | Yes | | Yes | Yes | |
| Unsuitability to the parties' interests | Yes | Yes | | Yes | No | | No | No | | Yes | Yes | | Yes | Yes | |
| Inadequate motivation | Yes | Yes | | Yes | No | | No | No | | Yes | Yes | | Yes | Yes | |
| **Evaluation criteria** | | | | | | | | | | | | | | | |
| Identification of the legally relevant institutes | 0 | 1 | **2** | 0 | 1 | **2** | 2 | 2 | **6** | 0 | 0 | **0** | 0 | 1 | **2** |
| Identification of the parties' interests | 0 | 1 | **2** | 0 | 1 | **2** | 2 | 2 | **6** | 0 | 0 | **0** | 0 | 1 | **2** |
| Adequacy of the selected legal institutes | 0 | 0 | **0** | 0 | 2 | **4** | 2 | 2 | **6** | 0 | 0 | **0** | 0 | 1 | **2** |
| Justification of that selection | 0 | 1 | **2** | 0 | 2 | **4** | 2 | 2 | **6** | 0 | 0 | **0** | 0 | 1 | **2** |
| Applicability of legal solution to the case | 0 | 1 | **2** | 0 | 2 | **4** | 2 | 2 | **6** | 0 | 0 | **0** | 0 | 0 | **0** |
| Quality of the deed's drafting | 0 | 0 | **0** | 0 | 2 | **4** | 2 | 2 | **6** | 0 | 0 | **0** | 0 | 0 | **0** |
| Discussion of the relevant legal institutes | 0 | 0 | **0** | 0 | 0 | **0** | 2 | 2 | **6** | 0 | 0 | **0** | 0 | 0 | **0** |
| **Total** | **0** | **8** | **8** | **0** | **20** | **20** | **14** | **28** | **42** | **0** | **0** | **0** | **0** | **8** | **8** |

Table 4: Grid for the mortis causa notary exam (consensus marks are weighted by number of concurring evaluators)